\documentclass[draftclsnofoot,onecolumn]{IEEEtran}
\usepackage{amsmath}
\usepackage{epsfig}
\usepackage{amssymb}
\usepackage{latexsym}
\usepackage{graphicx}
\usepackage{cite}

\title{Closed-Form Bounds to the Rice and \\ Incomplete Toronto Functions and \\ Incomplete Lipschitz-Hankel Integrals}
\author{P.~C.~Sofotasios,~\IEEEmembership{Member,~IEEE},
            and S.~Freear,~\IEEEmembership{Senior~Member,~IEEE}
\thanks{This work was carried out during the Doctoral studies of P. C. Sofotasios and was supported by the British Engineering and Physical Sciences Research Council (EPSRC)}
\thanks{P. C. Sofotasios and S. Freear are with the School of Electronic and Electrical Engineering, University of Leeds, LS2 9JT Leeds, UK \, (e-mail: $\lbrace$eenpso; s.freear$\rbrace$@leeds.ac.uk)}}
\begin{document}
\maketitle
\begin{abstract}
This article provides novel analytical results for the Rice function, the incomplete Toronto function and the incomplete Lipschitz-Hankel Integrals. Firstly, upper and lower bounds are derived for the Rice function, $Ie(k,x)$. Secondly, explicit expressions are derived for the incomplete Toronto function, $T_{B}(m,n,r)$, and the incomplete Lipschitz-Hankel Integrals of the modified Bessel function of the first kind, $Ie_{\mu,n}(a,z)$, for the case that $n$ is an odd multiple of $0.5$ and $m \geq n$. By exploiting these expressions, tight upper and lower bounds are subsequently proposed for both $T_{B}(m,n,r)$ function and $Ie_{\mu,n}(a,z)$ integrals. Importantly, all new representations are expressed in closed-form whilst the proposed bounds are shown to be rather tight. Based on these features, it is evident that the offered results can be utilized effectively in analytical studies related to wireless communications. Indicative applications include, among others, the performance evaluation of digital communications over fading channels and the information-theoretic analysis of multiple-input multiple-output systems.
\end{abstract}
\begin{keywords}
\noindent
Closed-form representations, Rice $Ie$-function, Incomplete Toronto function, Incomplete Lipschitz-Hankel Integrals, Marcum Q-function, lower and upper bounds
\end{keywords}
\section{Introduction}
\indent
It is widely accepted that special functions constitute invaluable mathematical tools in the majority of fields in natural sciences and engineering. In the area of telecommunications, their utilization in various studies studies often allows the derivation of  analytic expressions for important performance measures such as for example error probability and channel capacity. Furthermore, it has been shown that the computational realization special functions is not generally laborious since the most of them are included as built-in functions in popular scientific mathematical packages such as $Maple$, $Matlab$ and $Mathematica$. To this effect, both the algebraic representation and computation of any derived analytical expressions have been significantly simplified. \\
\indent
Among others, the Rice $Ie$-function, the incomplete Toronto function and the incomplete Lipschitz-Hankel integrals (ILHIs) appear in analytical solutions of numerous problems in telecommunications. They were all proposed several decades ago and they are denoted as $Ie(k,x)$, $T_{B}(m,n,r)$ and $Ze_{\mu,n}(a,z)$, respectively. \\
\indent
The Rice $Ie$-function was firstly proposed by S. O. Rice in [1] and has been largely exploited in the study of zero crossings, in the analysis of angle modulation systems, in radar pulse detection and in error rate analysis of differentially encoded systems, [2]${-}$[5]. Its definition is typically given in integral form which involves an exponential term and a modified Bessel function of the first kind and zeroth order. Alternative representations include two infinite series and one closed-form expression - reported in [4] and [5], respectively. The former, are expressed in terms of the modified Struve function and the modified Bessel function of the first kind whereas the latter is expressed in terms of the Marcum Q-function, $Q_{m}(a,b)$, [6]${-}$[10]. \\
\indent
In the same context, the incomplete Toronto function constitutes a special case of the Toronto function, which was initially proposed by Hatley in [11]. It also includes as a special case the Marcum Q-function and has been used in studies related to statistics, signal detection and estimation, radar systems and error probability analysis, [12]${-}$[14]. Its definition is also given in integral form while alternative representations include two infinite series, which were proposed in [15]. \\
\indent
Finally, the ILHIs are a class of incomplete cylindrical functions that have been largely encountered in analytical solutions of numerous problems in electromagnetics, [16]${-}$[17] and the references therein. In communication theory, they have been sufficiently utilized in investigations associated with the error rate analysis of MIMO systems under imperfect channel state information (CSI) that employ adaptive modulation, transmit beamforming and maximal ratio combining (MRC), [18]. \\ 
\indent
However, in spite of the evident importance of the $Ie(k,x)$ and $T_{B}(m,n,r)$ functions and the $Ze_{\mu,n}(a,z)$ integrals, it is noted that they are all neither tabulated, nor included as built-in functions in the aforementioned popular software packages. As a consequence, they are rather inconvenient to handle both analytically and computationally. \\
\indent
Motivated by these issues, this work is devoted in deriving novel representations for these functions and integrals. Specifically, upper and lower bounds to the $Ie(k,x)$ function and explicit expressions and upper and lower bounds to the $T_{B}(m,n,r)$ function and $Ie_{\mu,n}(a,z)$ integrals, are derived for the case that $n + 0.5 \in \mathbb{N}$ and $m \geq n$. Notably, the offered results are expressed in closed-form and have a tractable algebraic form. As a result, they can be meaningfully utilized in various analytical studies associated to wireless communications such as the performance evaluation of digital communications over fading channels and the information-theoretic analysis of MIMO systems. 
\section{Definitions and Existing Representations}
\subsection{The Rice $Ie$-function} 
$ $\\
\indent
The Rice $Ie$-function is defined as [4]${-}$[5],
\begin{equation} % Eq. 1
Ie(k,x) \triangleq \int_{0}^{x} e^{-t} I_{0}(kt) dt, \qquad 0 \leq k \leq 1
\end{equation}
where $I_{0}(.)$ is the modified Bessel function of the first kind and zero order, [6]${-}$[7]. An equivalent integral representation to $(1)$ was given in [4], namely,
\begin{equation} % Eq. 2
Ie(k,x) = \frac{1}{\sqrt{1 - k^{2}}} - \frac{1}{\pi} \int_{0}^{\pi} \frac{e^{-x(1 - cos\theta)}}{1 - k cos\theta} d\theta
\end{equation}
along with the following alternative series representations, 
$$
Ie(k,x) = \sqrt{\frac{x \pi}{2 \sqrt{1 - k^{2}}}} e^{-x} \sum_{n = 0}^{\infty} \frac{1}{n!} \frac{x^{n} k^{2n}}{2^{n} \sqrt{1 - k^{2}}} \times
$$
\begin{equation} % Eq. 3
\left[ \frac{1}{\sqrt{1 - k^{2}}} L_{n + \frac{1}{2}} \left(x\sqrt{1 - k^{2}} \right) + L_{n - \frac{1}{2}} \left(x\sqrt{1 - k^{2}} \right) \right]
\end{equation}
and
$$
Ie(k,x) = x e^{-x} \frac{\sqrt{\pi}}{2} \sum_{n = 0}^{\infty} \left( \frac{x\left(1 - k^{2} \right)}{2k} \right)^{n+1} \frac{I_{n+1}(kx)}{\Gamma \left(n + \frac{5}{2} \right)} \, +
$$
\begin{equation} % Eq. 4
x e^{-x} \left[I_{0}(kx) + \frac{\sqrt{\pi}}{2k} \sum_{n = 0}^{\infty} \left( \frac{x\left(1 - k^{2} \right)}{2k} \right)^{n} \frac{I_{n+1}(kx)}{\Gamma \left(n + \frac{3}{2} \right)} \right]
\end{equation}
The notations $L_{n}(.)$ and $\Gamma(.)$ denote the modified Struve function and the gamma function, respectively [6]${-}$[7]. Recalling [4], equation $(3)$ converges relatively quickly for the case that $x\sqrt{1 - k^{2}}$ is large and $kx$ is small. On the contrary, equation $(4)$ converges quickly when $x\sqrt{1 - k^{2}}$ is small and $kx$ is large. Nevertheless, this way of computation of $Ie(k,x)$ function is rather inefficient due to the following three facts: a) two relationships are required; b) the above series are relatively unstable due to their infinite form; c) the $L_{n}(.)$ function is not built-in in widely used mathematical software packages. \\
\indent
An adequate way of resolving this issue was reported in [5]. There, the Rice $Ie$-function is related to the Marcum Q-function of the first order, $Q_{1}(a,b)$ by the following relationships,
\begin{equation} % Eq. 5
Ie(k,x) = \frac{1}{\sqrt{1 - k^{2}}} \left[2Q(a , b) - e^{-x}I_{0}(kx) - 1 \right] 
\end{equation} 
and
\begin{equation} % Eq. 6
Ie(k,x) = \frac{1}{\sqrt{1 - k^{2}}} \left[ Q(a, b) - Q(b, a) \right]
\end{equation}
where, $a=\sqrt{x}\sqrt{1 + \sqrt{1 - k^{2}}}$ and $b=\sqrt{x}\sqrt{1 - \sqrt{1 - k^{2}}}$. \\
\subsection{The Incomplete Toronto Function}
$ $\\
\indent
The incomplete Toronto function is defined as,
\begin{equation} % Eq. 7
T_{B}(m,n,r) \triangleq 2r^{n-m+1} e^{-r^{2}} \int_{0}^{B} t^{m-n} e^{-t^{2}}I_{n}(2rt) dt
\end{equation}
Importantly, for the special case that $n = \frac{m-1}{2}$, it can be equivalently expressed in terms of the Marcum Q-function by the following relationship,
\begin{equation} % Eq. 8
T_{B}\left(m,\frac{m-1}{2},r\right) = 1 - Q_{\frac{m+1}{2}}\left(r\sqrt{2},B\sqrt{2}\right)
\end{equation}
Alternative representations to the $T_{B}(a,b,r)$ function, in the form of infinite series, were reported in [15], namely, 
\begin{equation} % Eq. 9
T_{B}(m,n,r) = \frac{B^{2a}r^{2(n-a+1)}}{n!}e^{-B^{2}-r^{2}}\sum_{k=0}^{\infty}\frac{r^{2k}Y_{k}}{(a)_{k+1}}
\end{equation}
and 
\begin{equation} % Eq. 10
T_{B}(m,n,r) = r^{2(n-a+1)}e^{-r^{2}}\sum_{k=0}^{\infty} \frac{r^{2k}\gamma(a+k, B)}{k!(n+k)!}
\end{equation}
where the notations $(a)_{k}$ and $\gamma(c,x)$ denote the Pochhammer symbol and the lower incomplete gamma function, respectively [6]${-}$[7]. Also, 
$$
Y_{k} = \sum_{i=0}^{k}\frac{(a)_{i}r^{2i}}{(n+1)_{i} i!} 
$$
and
$$
a = \frac{m+1}{2} 
$$
Although equation $(9)$ is exact, its algebraic representation is rather inconvenient both analytically and numerically. Equation $(10)$ is significantly more tractable than $(9)$, yet its infinite form eventually raises convergence and truncation issues. 
\subsection{The Incomplete Lipschitz-Hankel Integrals}
$ $\\
\indent
The general ILHI is defined as, 
\begin{equation} % Eq. 11
Ze_{m,n}(x;a) \triangleq \int_{0}^{z}x^{m}e^{-ax}Z_{n}(x)dx
\end{equation}
where $m$,$n$,$a$,$z$ may be complex [16]-[17]. The notation $Z_{n}(x)$ denotes one of the cylindrical functions\footnote{Only the $I_{n}(x)$ function is considered in the present analysis.} $J_{n}(x)$, $I_{n}(x)$, $Y_{n}(x)$, $K_{n}(x)$, $H_{n}^{1}(x)$ or $H_{n}^{2}(x)$, [6]-[7]. An alternative representation for the ILHIs of the first-kind modified Bessel functions, was recently reported in [18]. This representation is given in terms of the Marcum Q-function and is expressed as follows, 
$$
Ie_{m,n}(x;a) = A_{m,n}^{0}(a) + e^{-ax} \sum_{i=0}^{m}\sum_{j=0}^{n+1}B_{m,n}^{i,j}(a)x^{i}I_{j}(x) 
$$
\begin{equation} % Eq. 12 
\qquad \, \qquad \, \qquad \, \qquad \, \quad + A_{m,n}^{1}(a)Q_{1}\left(\sqrt{\frac{x}{a+\sqrt{a^{2}-1}}}, \sqrt{x}\sqrt{a+\sqrt{a^{2}-1}} \right)
\end{equation}
where the set of coefficients $A_{m,n}^{l}(a)$ and $B_{m,n}^{i,j}(a)$ can be obtained recursively, [18]. As aforementioned, the above relationship was found useful in analytical investigations related to error rate of MIMO systems under imperfect channel state information (CSI). \\
\section{Performance Bounds to the Rice $Ie$-function}
$ $\\
\indent 
This section is devoted to the derivation of upper and lower bounds for the Rice $Ie$-function. To this end, it is a critical to express $Ie(k,x)$ function alternatively. 
\subsection{An Alternative Representation to the $Ie(k,x)$ Function}
$ $\\
\noindent
\textbf{Lemma 1.}\textit{ For $x>0$ and $0 \leq k \leq 1$, the following relationship holds,} 
$ $\\
\begin{equation} % Eq. 13
Ie(k,x) = 1 - e^{-x}I_{0}(kx) + k \int_{0}^{x} e^{-t} I_{1}(kt)dt
\end{equation}
$ $\\
\noindent
\textit{Proof.} By integrating once equation $(1)$ by part, it follows that,
\begin{equation} % Eq. 14
Ie(k,x) = \left[ \int_{0}^{x} e^{-t} dt \right] I_{0}(kt) - \int_{0}^{x} \left[\int_{0}^{x} e^{-t} dt \right]\left[\frac{d}{dt} I_{0}(kt) \right] dt
\end{equation}
According to the basic principles of integration, the first integral yields straightforwardly,
\begin{equation} % Eq. 15
\int_{0}^{x} e^{-t} dt = -e^{-x}
\end{equation}
Subsequently, based on [6]${-}$[7], the derivative of the modified Bessel function of the first kind and order $n$ is re-written as,
\begin{equation} % Eq. 16
\frac{d}{dx} I_{n}(kx) =\frac{k}{2} \left[ I_{n-1}(kx) + I_{n+1}(kx) \right]
\end{equation}
which for $n=0$ reduces to
\begin{equation} % Eq. 17
\frac{d}{dt} I_{0}(kt) = kI_{1}(kt)
\end{equation}
Consequently, by setting $x=t$, substituting $(15)$ and $(17)$ into $(14)$ and noticing that $I_{n}(0) = 1$, equation $(13)$ is deduced. $\blacksquare$ 
\subsection{An Upper Bound to the $Ie(k,x)$ Function}
$ $\\
\noindent
\textbf{Theorem 1.} \textit{For $x>0$ and $0 \leq k \leq 1$, the following inequality holds,}
$ $\\
\begin{equation} % Eq. 18
Ie(k,x) < 1 - e^{-x} I_{0}(kx) + \sqrt{\frac{k}{2}} \left[ \frac{erf(\sqrt{x}\sqrt{1 - k})}{\sqrt{1 - k}} - \frac{erf( \sqrt{x}\sqrt{1 + k})}{\sqrt{1 + k}} \right]
\end{equation}
$ $\\
\noindent
\textit{Proof.} It is recalled here that the modified Bessel function of the first kind is strictly decreasing with respect to its order, $n$. Therefore, for $a>0$ it immediately follows that $I_{n-a}(x) > I_{n}(x)$. Thus, for $n \pm a \pm \frac{1}{2} \in \mathbb{N} $, one obtains the following inequality, 
\begin{equation} % Eq. 19
Ie(k,x) < 1 - e^{-x}I_{0}(kx) + k \int_{0}^{x} e^{-t} I_{\frac{1}{2}}(kt)dt 
\end{equation}
Notably, for the special case that $n$ is an odd multiple of $0.5$, i.e. $n +0.5 \in \mathbb{N}$, a closed-form representation for the $I_{n}(x)$ function is given in $[6, eq.(8.467)]$, namely,
\begin{equation} % Eq. 20
I_{n + \frac{1}{2}}(x)\triangleq \sum_{k=0}^{n}\frac{(n+k)!\,\left[(-1)^{k}e^{x} + (-1)^{n+1}e^{-x}\right]}{\sqrt{\pi}k!(n-k)!(2x)^{k+\frac{1}{2}}}, \, \qquad \, n\in \mathbb{N}
\end{equation}
Therefore, for $n=0$, eq.$(20)$ reduces to,
\begin{equation} % Eq. 21
I_{\frac{1}{2}}(kt) = \frac{e^{kt} - e^{-kt}}{\sqrt{2\pi kt}} 
\end{equation}
Evidently, the proof of the theorem is subject to evaluation of the integral in $(19)$. To this end, by substituting $(21)$ into $(19)$, one obtains,
\begin{equation} % Eq. 22
\int_{0}^{x} e^{-t} I_{\frac{1}{2}}(kt)dt = \int_{0}^{x} e^{-t} \left[\frac{e^{kt} - e^{-kt}}{\sqrt{2\pi kt}}\right] dt 
\end{equation}
which has the following closed-form solution,
\begin{equation} % Eq. 23
\int_{0}^{x} e^{-t} I_{\frac{1}{2}}(kt)dt = \sqrt{\frac{k}{2}} \left[ \frac{erf(\sqrt{x}\sqrt{1 - k})}{\sqrt{1 - k}} - \frac{erf(\sqrt{x} \sqrt{1 + k})}{\sqrt{1 + k}} \right]
\end{equation}
where,
$$
erf(x) \triangleq \frac{2}{\sqrt{\pi}} \int_{0}^{x} e^{-t^{2}} dt
$$
is the error function, [6]${-}$[7]. Finally, by substituting $(23)$ into $(19)$, the proof is completed. $\blacksquare$ 
\\
$ $\\
\noindent
\textit{Remark.} The authors in [19] derived closed-form bounds to the Marcum Q-function. Thus, by making the necessary change of variables and make the according substitution in equations $(5)$ and/or $(6)$, an alternative expression to $(18)$ can be deduced. However, this expression is significantly less compact and convenient than $(18)$ both analytically and numerically. 
\subsection{A Lower Bound to the $Ie(k,x)$ Function} 
$ $ \\
\noindent
\textbf{Theorem 2.} \textit{For $x>0$ and $0 \leq k \leq 1$, the following inequality holds,}
$ $
\\
\begin{equation} % Eq. 24
Ie(k,x)>\frac{2Q(b+a)+2Q(b-a) - e^{-x} I_{0}(kx) - 1}{\sqrt{1-k^{2}}}
\end{equation}
where $Q(.)$ is the one dimensional Gaussian Q-function [6],
$$
Q(x) \triangleq \frac{1}{\sqrt{2 \pi}} \int_{x}^{\infty} e^{-\frac{t^{2}}{2}} dt
$$
$ $\\
\noindent
\textit{Proof.} According to the aforementioned monotonicity property of the $I_{n}(x)$ function it follows that $I_{\frac{3}{2}}(x) < I_{1}(x)$. Therefore, by making the necessary change of variables and substituting in $(13)$, one obtains,
\begin{equation} % Eq. 25
Ie(k,x) > 1 - e^{-x}I_{0}(kx) + k \int_{0}^{x} e^{-t} I_{\frac{3}{2}}(kt)dt 
\end{equation}
Importantly, a similar inequality may be also deduced by exploiting equations $(5)$ and $(6)$. To this end, it is firstly recalled that the Marcum Q-function is strictly increasing with respect to its order $m$. Based on this, it follows that
\begin{equation} % Eq. 26
Q_{1}(a,b)>Q_{\frac{1}{2}}(a,b)
\end{equation}
Subsequently, by substituting $(26)$ into $(5)$, the following inequality is deduced,
\begin{equation} % Eq. 27
Ie(k,x) > \frac{1}{\sqrt{1 - k^{2}}} \left[2Q_{\frac{1}{2}}(a, b) - e^{-x}I_{0}(kx) - 1 \right]
\end{equation}
Of note, the authors in [19] show that, 
\begin{equation} % Eq. 28
Q_{\frac{1}{2}}(a,b) = Q(b+a) + Q(b-a)
\end{equation} 
Therefore, by substituting $(28)$ into $(27)$, eq.$(19)$ is obtained and thus the proof is completed. $\blacksquare$\\
$ $
\\
\textit{Remark.} A lower bound to $Ie(k,x)$ could be theoretically derived by following the same methodology as in Theorem $1$. Nevertheless, this approach ultimately renders a representation that is both complex and divergent. \\
\subsection{Numerical Results}
$ $ \\
\indent
The behaviour and tightness of the derived bounds is illustrated in figures $1$-$4$. In more details, figure $1$ depicts the bounds in $(18)$ and $(24)$ for $k=0.5$ along with results obtained from numerical integrations for comparative purposes. Evidently, the upper bound is tighter than the lower bound for small values of $x$. However, its tightness degrades as $x$ increases while the lower bound becomes much tighter. According to figure $2$, this appears to be also the case for a relatively small $x$ and different values of $k$. \\
\indent
Interestingly enough, for large values of $x$- typically $x>40$- the lower bound in $(24)$ is so tight that it eventually becomes a highly accurate approximation to $Ie(k,x)$. This is clearly illustrated in figure $3$ where for $x=80$ and different $k$, the plotted curve provides an excellent match to the corresponding theoretical results. This is also justified by the overall small absolute relative error, $\epsilon_{ar} = \mid Ie(k,x) - eq.(19) \mid/ Ie(k,x)$, which is $\epsilon_{ar} < 10^{-6}$ over almost the whole range of values of $k$. \\
\section{A Closed-form Representation and Bounds to the Incomplete Toronto Function}
\indent 
As mentioned in section $I$, the $T_{B}(m,n,r)$ is neither expressed in terms of other elementary and/or special functions, nor is a built-in function in popular mathematical software packages. Motivated by this, we derive a closed-form representation for the case that $n$ is an odd multiple of $0.5$. Subsequently, we exploit this representation to propose novel closed-form upper and lower bounds. 
\subsection{A Closed-form Solution to the $T_{B}(m,n,r)$ Function}
\noindent
\textbf{Theorem 3.} \textit{For $m,r,B \geq 0$, $m\geq n + \frac{1}{2}$ and $n + 0.5 \in \mathbb{N}$, the following closed form relationship holds,}
$ $\\
\begin{equation} % Eq. 29
T_{B}(m,n,r) = \sum_{k=0}^{n-\frac{1}{2}} \sum_{l=0}^{L} \frac{\left(n+k-\frac{1}{2}\right)!\,L!\,2^{-2k-1}r^{-2k-l}}{\sqrt{\pi}k!\left(n-k-\frac{1}{2} \right)!\,l!\, (L-l)!} \times
\end{equation}
$$
\left\lbrace (-1)^{m-k-l} \left[\Gamma \left(\frac{l+1}{2}, r^{2} \right) - \Gamma \left(\frac{l+1}{2}, (n+r)^{2} \right)\right] + (-1)^{k} \gamma\left(\frac{l+1}{2}, (b-r)^{2} \right) + (-1)^{k+l} \gamma \left(\frac{l+1}{2}, r^{2} \right) \right\rbrace
$$
where $L = m-n-k-\frac{1}{2}$ and $\Gamma(a,x)$, $\gamma(a,x)$ denote the upper and lower incomplete gamma functions, respectively. 
\\
$ $ \\
\noindent
\textit{Proof.} By setting in $(20)$ $x = 2rt$, the $I_{n}(x)$ function is re-written as,
\begin{equation} % 30
I_{n}(2rt) = \frac{1}{\sqrt{\pi}} \sum_{k=0}^{n - \frac{1}{2}} \frac{\left(n+k+\frac{1}{2} \right)! r^{-k - \frac{1}{2}}}{k!\left(n - k - \frac{1}{2} \right)! 2^{2k+1}} \left[ \frac{(-1)^{k} e^{2rt} + (-1)^{n+\frac{1}{2}}e^{-2rt}}{t^{k+\frac{1}{2}}}\right], \, \quad \, n + \frac{1}{2} \in \mathbb{N}
\end{equation}
By substituting into $(7)$ and utilizing the basic identity $(a \pm b)^{2} = a^{2} \pm 2ab +b^{2}$, it follows that,
$$
T_{B}(m,n,r) = \sum_{k=0}^{n - \frac{1}{2}} \frac{(n+k - \frac{1}{2})! r^{n-m-k-\frac{1}{2}}}{k!\sqrt{\pi}(n-k-\frac{1}{2})!2^{2k}} \times
$$
\begin{equation} % 31
\left[(-1)^{k} \int_{0}^{B} t^{L} e^{-(t-r)^{2}}dt + (-1)^{n+\frac{1}{2}} \int_{0}^{B} t^{L} e^{-(t+r)^{2}} dt \right]
\end{equation}
where $L=m-n-k-\frac{1}{2}$. Evidently, a closed-form solution to the above expression is subject to evaluation of the two involved integrals. To this end, with the aid of $[21, eq.(1.3.3.18)]$, the above expression is re-written equivalently as follows,
$$
T_{B}(m,n,r) = \sum_{k=0}^{n - \frac{1}{2}} \sum_{l=0}^{L}\frac{\left(n+k-\frac{1}{2} \right)!\, L!\,2^{-2k}r^{-2k-l}}{\sqrt{\pi}k! \left(n-k-\frac{1}{2} \right)!l!(L-l)!} \times
$$
\begin{equation} % 32 
\left[(-1)^{m-k-l}\int_{r}^{\infty}t^{l}e^{-t^{2}}dt - (-1)^{m-k-l}\int_{B+r}^{\infty}t^{l}e^{-t^{2}}dt - (-1)^{k} \int_{B-r}^{\infty} t^{l}e^{-t^{2}}dt + (-1)^{k}\int_{-r}^{\infty} t^{l}e^{-t^{2}}dt \right]
\end{equation}
Finally, with the aid of $[6, eq.(3.381.3)]$, eq.$(29)$ is deduced and thus, the proof is completed. $\blacksquare$ 
\subsection{Upper and lower bounds to the $T_{B}(m,n,r)$ function}
$ $ \\
\indent
With the aid of Theorem $3$, explicit bounds to the incomplete Toronto function may be straightforwardly deduced. \\
$ $\\
\noindent
\textbf{Corollary 1.} \textit{For $m,r,B \geq 0$ and $n \in \mathbb{N}$, $m\geq n$, the following inequality holds,}
\begin{equation} % Eq. 33
T_{B}(m,n,r) > T_{B}\left(m,n + \frac{1}{2}, r \right)
\end{equation}
where $T_{B}\left(m,n + \frac{1}{2}, r \right)$ is given in closed-form in eq.$(29)$.
\\
$ $ \\
\noindent
\textit{Proof.} The incomplete Toronto function is strictly decreasing with respect to $n$. Therefore, for an arbitrary real positive value $a$, it follows that $T_{B}(m,n+a,r)<T_{B}(m,n,r)$. Hence, for the case that $n \in \mathbb{N}$ and $a=0.5$, the lower bound in $(33)$ is straightforwardly deduced in closed-form. $\blacksquare$\\
$ $\\
\noindent
\textbf{Corollary 2.} \textit{For $m,r,B \geq 0$ and $n \in \mathbb{N}$, $m\geq n$, the following inequality holds,}
\begin{equation} % Eq. 34
T_{B}(m,n,r) < T_{B}\left(m,n - \frac{1}{2}, r \right)
\end{equation}
where $T_{B}\left(m,n - \frac{1}{2}, r \right)$ is given in closed-form in eq.$(29)$.
\\
$ $ \\
\noindent
\textit{Proof.} The proof follows immediately from Theorem $3$ and Corollary $1$. $\blacksquare$ 
\subsection{Numerical Results} 
$ $ \\
\indent
The validity of the closed-form expression and the behaviour and tightness of the derived bounds are shown in figures $4$ and $5$ with respect to $r$. In the former, eq.$(29)$ is depicted for $n=0.5$ and $m=1.0$ along with results obtained from numerical integrations for $m=1.0$ and $n=0.4$, $n=0.5$ and $n=0.6$. In the latter, eq.$(29)$ is depicted for $m=3.0$ and $n=2.5$ along with numerical results for $n=2.4$, $n=2.5$ and $2.6$. It is evident that $(29)$ is in exact agreement with the numerical results while the overall tightness of the derived bounds is quite adequate over the whole range of values of $r$. 
\section{A closed-form representation and performance bounds to the incomplete Lipschitz-Hankel integrals}
$ $\\
\indent 
Likewise the $T_{B}(m,n,r)$ function, the ILHIs are neither tabulated, nor built-in in popular mathematical software packages. However, their algebraic form renders possible the derivation of a closed-form expression for the case that $n$ is an odd multiple of $0.5$. By exploiting this result, upper and lower bounds are deduced based on the same criteria as in the previous section.\\
\subsection{A Closed-Form Solution to the $Ie_{m,n}(z;a)$ Integrals}
$ $ \\
\noindent
\textbf{Theorem 4.} \textit{For $m\geq n$, $m,r,B \geq 0$ and $n + 0.5 \in \mathbb{N}$, the following closed-form relationship holds,}
$ $\\
\begin{equation} % Eq. 35
Ie_{m,n}(z;a) = \sum_{k=0}^{n - \frac{1}{2}} \frac{\left(n+k-\frac{1}{2} \right)!}{\sqrt{\pi}k!\left(n-k-\frac{1}{2} \right)!2^{k+\frac{1}{2}}} \left[(-1)^{k} \frac{\gamma \left(P, (a-1)z \right) }{(a-1)^{P}} + (-1)^{n + \frac{1}{2}} \frac{\gamma \left(P, (a+1)z \right) }{(a+1)^{P}} \right]
\end{equation}
where $P=m-k+\frac{1}{2}$. 
\\
$ $ \\
\noindent
\textit{Proof.} By substituting the $I_{n}(x)$ function with its closed-form representation in $(20)$, equation $(11)$ can be equivalently re-written as,
\begin{equation} % 36
Ie_{m,n}(z;a) = \sum_{k=0}^{n-\frac{1}{2}} \frac{\left(n+k- \frac{1}{2}\right)! 2^{-k - \frac{1}{2}}}{\sqrt{\pi} k!\left(n-k-\frac{1}{2} \right)!} \left[(-1)^{k}\int_{0}^{z}x^{P}e^{-ax}e^{x}dx + (-1)^{n + \frac{1}{2}} \int_{0}^{z} x^{P}e^{-ax}e^{-x}dx \right]
\end{equation}
Notably, the integrals in $(36)$ clearly belong to the family of gamma special functions. Therefore, after some basic algebraic manipulation and with the aid of $[6, eq.(3.381.3)]$, eq.$(35)$ is deduced. $\blacksquare$ \\
$ $\\
\noindent
\textit{Remark.} The present analysis was limited in the consideration of only the $I_{n}(x)$ function in $(11)$. However, similar results may be also derived analogously for the Bessel functions $J_{n}(x)$, $Y_{n}(x)$, $K_{n}(x)$ and the Hankel functions $H_{n}^{(1)}(x)$, $H_{n}^{(2)}(x)$. 
\subsection{Upper and Lower Bounds to the $Ie_{m,n}(z;a)$ Integrals}
$ $ \\
\noindent
\textbf{Corollary 3.} \textit{For $m\geq n$, $m,r,B \geq 0$ and $n \in \mathbb{N}$, the following inequality holds}
\begin{equation} % Eq. 37
I_{m,n}(z;a) > I_{m,n + \frac{1}{2}}(z;a)
\end{equation}
where the $I_{m,n + \frac{1}{2}}(z;a)$ is given in $(35)$ in closed-form.
\\
$ $ \\
\noindent
\textit{Proof.} The $I_{m,n}(z;a)$ integrals are strictly decreasing with respect to $n$. Therefore, for an arbitrary real positive value $a$, it follows that $I_{m+a,n}(z;a) < I_{m,n}(z;a)$. Thus, for the case that $n \in \mathbb{N}$ and $a=0.5$, one obtains the the closed-form lower bound in $(35)$. $\blacksquare$\\
$ $\\
\noindent
\textbf{Corollary 4.} \textit{For $m\geq n$, $m,r,B \geq 0$ and $n \in \mathbb{N}$, the following inequality holds}
\begin{equation} % Eq. 38
I_{m,n}(z;a) < I_{m,n - \frac{1}{2}}(z;a)
\end{equation}
where $I_{m,n - \frac{1}{2}}(z;a)$ is given in $(33)$ in closed-form.
\\
$ $ \\
\noindent
\textit{Proof.} The proof follows immediately from Theorem $4$ and Corollary $3$. $\blacksquare$ \\
\subsection{Numerical Results}
$ $ \\
The validity and behaviour of the offered results are explicitly illustrated in figure $6$. Specifically, one can observe the exactness of $(35)$ and the rather tight performance of the performance bounds in all areas of values of their arguments. \\
\section{Conclusion}
\indent 
In this work, explicit representations and performance bounds for the Rice $Ie$-function, the incomplete Toronto function and the incomplete Lipschitz-Hankel integrals of the modified Bessel function of the first kind were derived. The offered results are novel and are all expressed in closed-form. This property is sufficiently advantageous since it renders them suitable for efficient utilization in various analytical applications in the wide area of digital communications. 
\bibliographystyle{IEEEtran}
\begin{Large}
\begin{center}
\underline{LIST OF TABLES AND FIGURES}
\end{center}
\end{Large}
$$ $$
$$ $$
Figure 1: Behaviour of the bounds in equations $(18)$ and $(24)$ for $k=0.5$
\\\\
Figure 2: Behaviour of the bounds in equations $(18)$ and $(24)$ for $x=7.0$
\\\\
Figure 3: Behaviour of the lower bound in $(24)$ for $x=40$ 
\\\\
Figure 4: Behaviour of the exact solution and the performance bounds to $T_B(m,n,r)$ for $m=1.0$ and different $n$
\\\\
Figure 5: Behaviour of the exact solution and the performance bounds to $T_B(m,n,r)$ for $m=3.0$ and different $n$
\\\\
Figure 6: Behaviour of the exact solution and the performance bounds to $Ie_{m,n}(z;a)$ for different values of $n$ and $m$
\newpage
\begin{figure}
\includegraphics{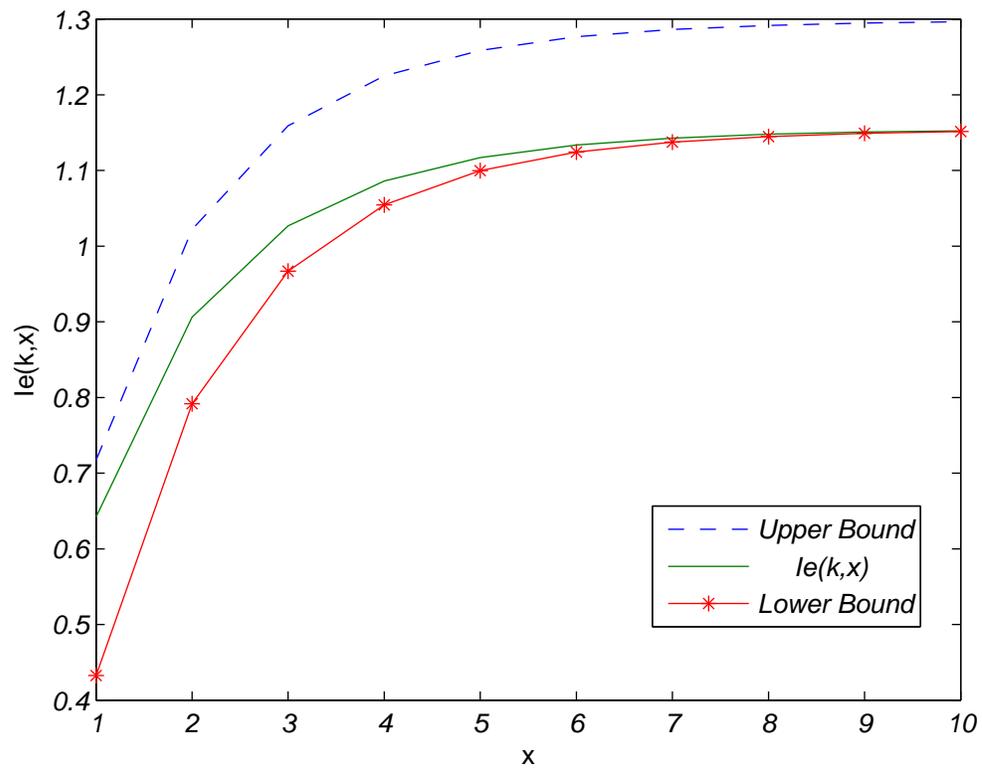} 
\caption{Behaviour of the bounds in equations $(18)$ and $(24)$ for $k=0.5$}
\end{figure}
\begin{figure}
\includegraphics{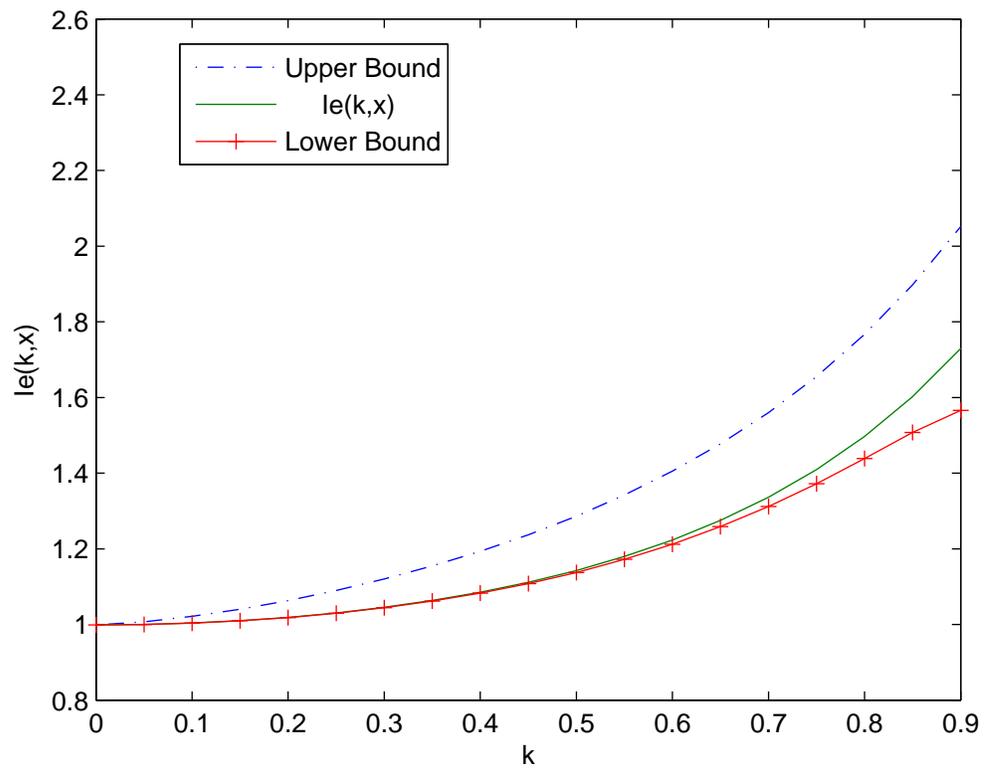} 
\caption{Behaviour of the bounds in equations $(18)$ and $(24)$ for $x=7.0$}
\end{figure}
\begin{figure}
\includegraphics{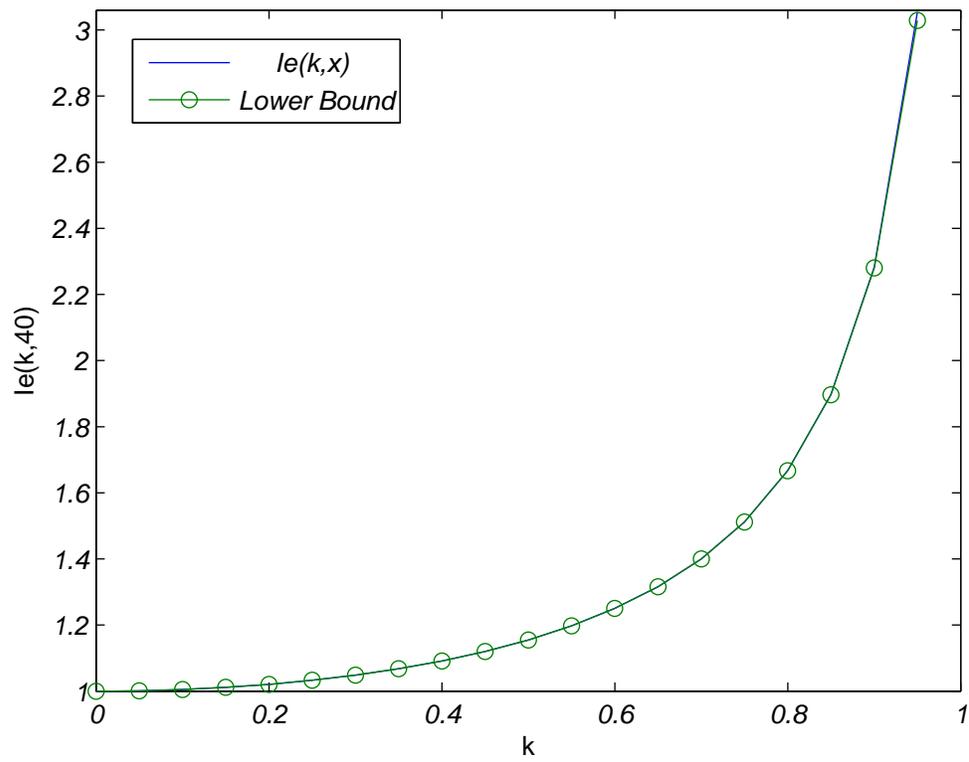} 
\caption{Behaviour of the lower bound in $(24)$ for $x=40$}
\end{figure}
\begin{figure}
\includegraphics{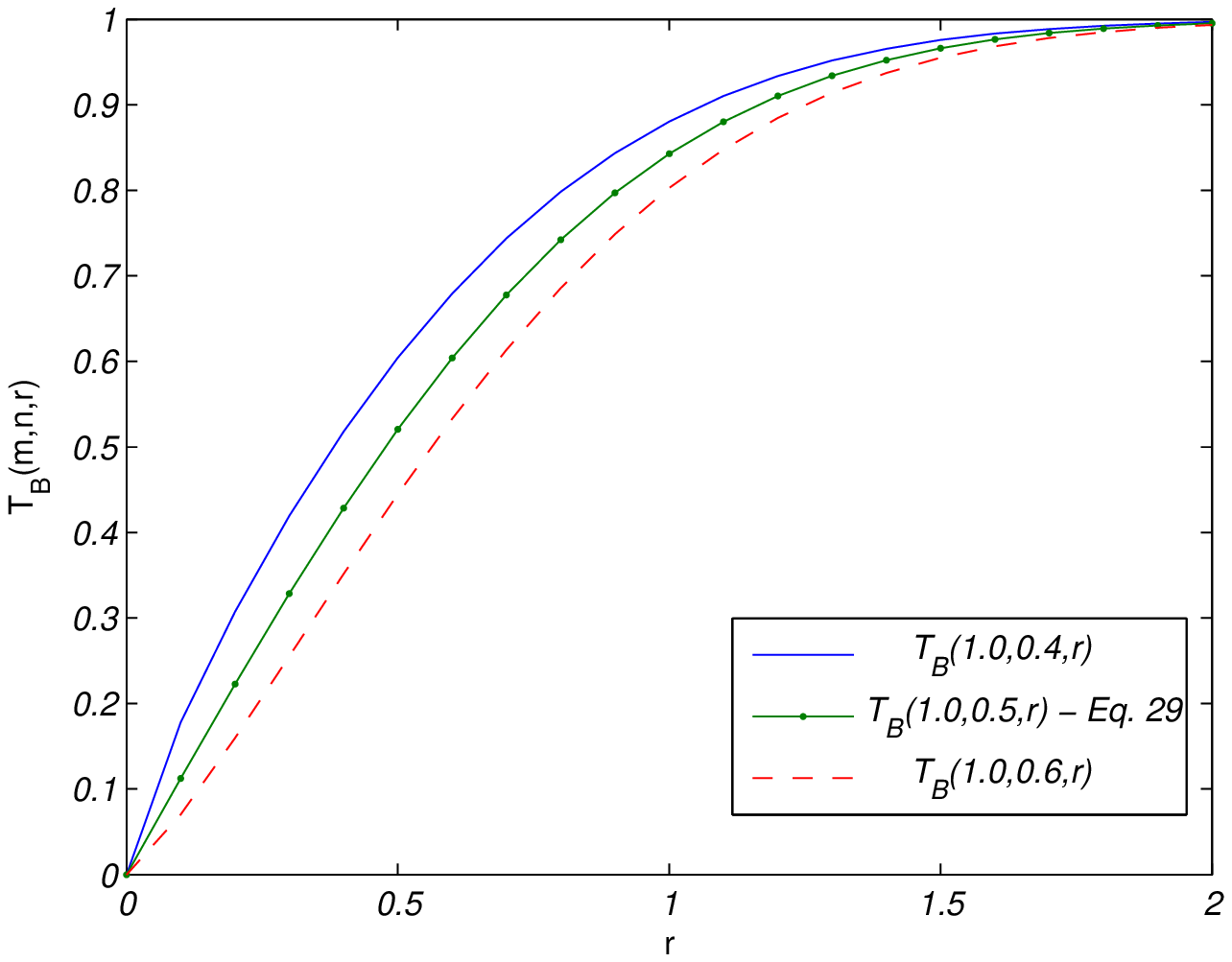} 
\caption{Behaviour of the exact solution and the performance bounds to $T_B(m,n,r)$ for $m=1.0$ and different $n$}
\end{figure}
\begin{figure}
\includegraphics{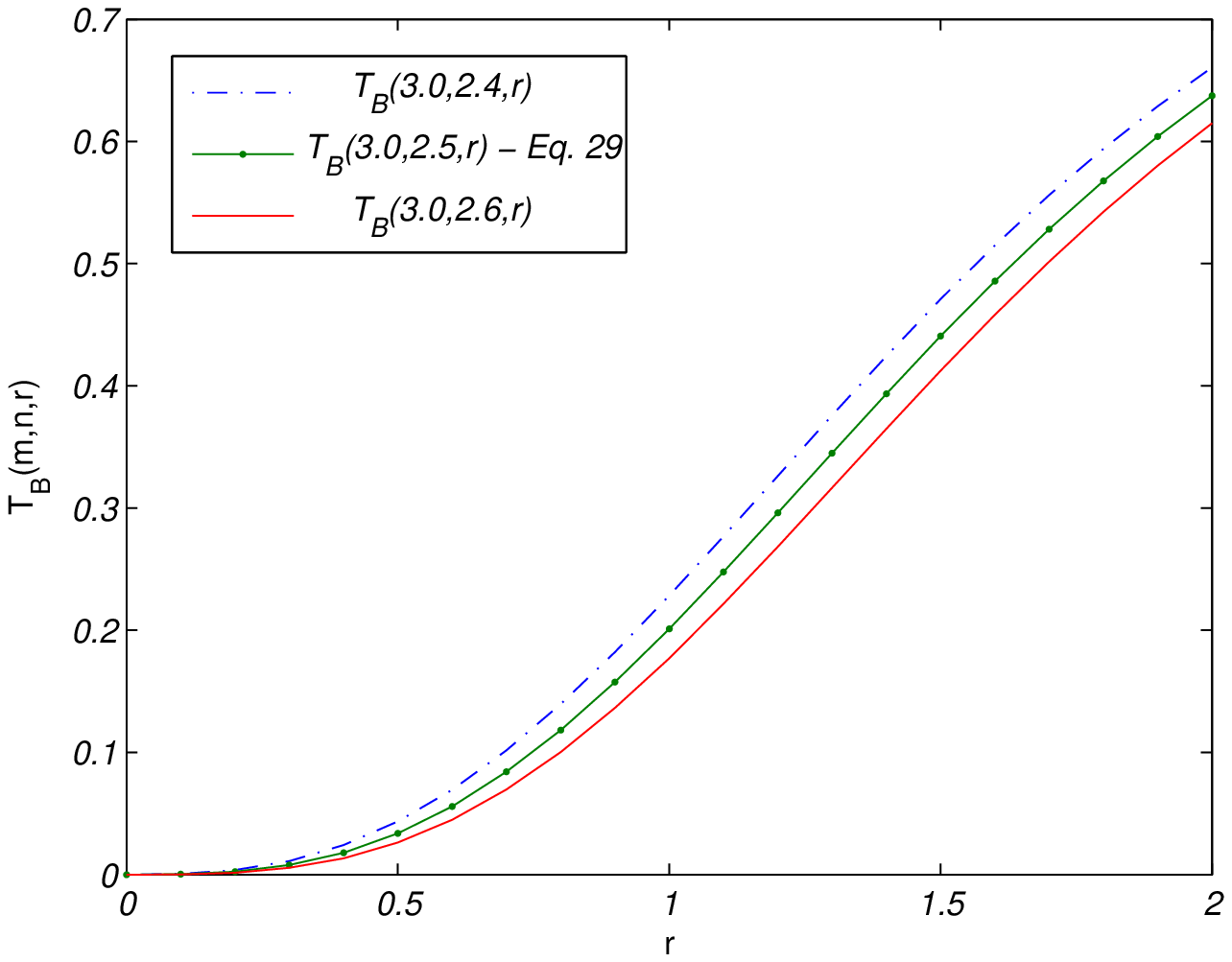} 
\caption{Behaviour of the exact solution and the performance bounds to $T_B(m,n,r)$ for $m=3.0$ and different $n$}
\end{figure}
\begin{figure}
\includegraphics{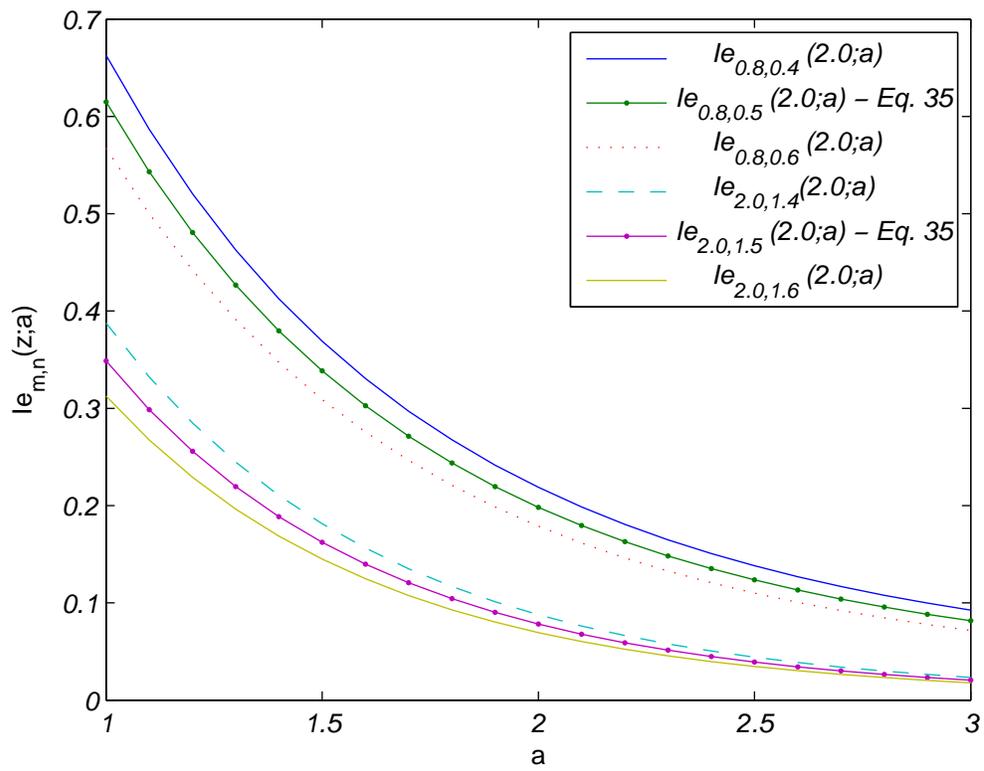} 
\caption{Behaviour of the exact solution and the performance bounds to $Ie_{m,n}(z;a)$ for different values of $n$ and $m$}
\end{figure}
\end{document}